\documentclass[showpacs,preprintnumbers,amsmath,amssymb]{revtex4}

\usepackage{graphicx}
\usepackage{amsfonts}
\usepackage{amssymb}
\usepackage{latexsym}
\usepackage{cancel}
\usepackage{color}

 \newcommand{\CL}{{\cal L}}

\newcommand{\bear}{\begin{array}}  \newcommand{\eear}{\end{array}}
\newcommand{\bea}{\begin{eqnarray}}  \newcommand{\eea}{\end{eqnarray}}
\newcommand{\beq}{\begin{equation}}  \newcommand{\eeq}{\end{equation}}
\newcommand{\bef}{\begin{figure}}  \newcommand{\eef}{\end{figure}}
\newcommand{\bec}{\begin{center}}  \newcommand{\eec}{\end{center}}
  
\newcommand{\lmk}{\left(}  \newcommand{\rmk}{\right)}

\newcommand{\vecs}[1]{\mbox{\scriptsize\boldmath${#1}$}}

\newcommand{\order}{{\cal O}}

\newcommand{\rr}{B}
\newcommand{\calrk}{{\cal R}_{\vecs k}}

\def\apjs #1 #2 #3 {#1 ApJS, {\bf #2} #3}
\def\aap  #1 #2 #3 {#1 A\&A, {\bf #2} #3}
\def\mnras #1 #2 #3 {#1 MNRAS, {\bf #2} #3}
\def\physrep #1 #2 #3 {#1 Phys. Rep., {\bf #2} #3}
\def\jcap #1 #2 #3{J. Cosmol. Astropart. Phys. {\bf #1}, #2 #3}

\begin{document}

\title{Cosmic Discordance: Detection of a modulation in the primordial
fluctuation spectrum}

\author{Kiyotomo Ichiki$^{1,2}$,~Ryo Nagata$^{1,3}$,~and~Jun'ichi
Yokoyama$^{1,4}$}
\email{yokoyama@resceu.s.u-tokyo.ac.jp}
\affiliation{%
$^1$Research Center for the Early Universe, University of Tokyo, 7-3-1
Hongo, Bunkyo-ku, Tokyo 113-0033, Japan
}
\affiliation{%
$^2$Department of Physics and Astrophysics, Nagoya
University, Nagoya 464-8602, Japan
}
\affiliation{%
$^3$Cosmophysics Group, IPNS, KEK, Tsukuba 305-0801, Japan
}
\affiliation{%
$^4$Institute for the Physics and Mathematics of the
Universe,\\  The University of Tokyo, Kashiwa 277-8568, Japan
}

\preprint{RESCEU-29/09}

\begin{abstract}
As a test of the standard inflationary cosmology,
which generically predicts nearly scale-invariant spectrum
of primordial curvature fluctuations, we  perform Markov-Chain 
Monte-Carlo analysis to search for possible  modulations 
in the power spectrum and determine its shape together with the 
cosmological parameters using cosmic microwave background radiation
 data.  By incorporating various
three-parameter features on the simple power-law spectrum, we 
find an oscillatory modulation localized around the comoving wavenumber 
$k\simeq 0.009\mathrm{Mpc}^{-1}$ at 99.995\% confidence level which 
improves the log-likelihood as much as  
$-\Delta 2\ln {\cal L}\equiv \Delta \chi^2_{\rm eff}=-22$.  
This feature can be detected even if we use only the cross 
correlation between the temperature and the E-mode polarization 
anisotropies.
\end{abstract}


\maketitle

\section{Introduction}

In the standard inflationary cosmology \citep{guth,sato,staro,infreview} 
the observed large-scale structures and the anisotropies in the 
cosmic microwave background radiation (CMB)
originate in tiny quantum fluctuations generated during the 
inflationary expansion stage in the very early
 Universe \citep{yuragi,yuragi2,yuragi3}.  
Conventionally, the power spectrum of primordial curvature fluctuations,
$P(k)\equiv \langle |\calrk|^2\rangle$, has been 
assumed to follow
a simple functional form such as a power-law  $A(k)\equiv 
\frac{4\pi k^3}{(2\pi)^3}P(k)=A(k/k_0)^{n_s-1}$ 
which is quantified by the amplitude $A$ and the spectral index $n_s$,
and the previous statistical analyses of the power spectrum have 
mostly focused on these two parameters (and at best the scale dependence
of $n_s$, the running, which may be important to distinguish 
the inflation
model\citep{Kawasaki:2003zv}).  It is true that the simplest class
of inflation models 
predicts a  power-law spectrum with
$n_s$ close to unity \citep{yuragi,yuragi2,yuragi3}, 
but we may not be able to 
identify the correct theory of the early Universe if we restrict
our parameter space from the beginning.  From the viewpoint of
observational cosmology, the shape of the primordial fluctuation 
spectrum should be determined purely from the observational data
without any theoretical prejudices.

Along this line of thought, several Markov-Chain 
Monte-Carlo (MCMC) analyses have been  
performed to 
search for possible deviation from the power law, but none 
has detected
statistically much significant modulations so 
far \citep{4,4.2,5,7,WMAP3COSMO,2009PhRvD..80h3002I}.  
To be more quantitative,
the presence of a nontrivial feature may be decided in terms of 
Akaike's Information Criterion (AIC) \citep{akaike}
which asserts that introduction of an
additional fitting parameter is justified if and only if
$\chi^2_\mathrm{eff}$ improves more than $-2$ with it.
Most of the previous analyses resulted in the improvement of 
$\Delta\chi^2_\mathrm{eff}$
much less than that required by AIC, and others reached
it only at a marginal level, so that it was hardly possible to 
claim the presence of a feature.
It should be emphasized, however, that those analyses 
did not have sufficient resolution to detect spectral fine 
structure,  because they
tried to fit the primordial power spectrum in a broad range of scales
in terms of a limited number of degrees of freedom
using sparse sampling or some specific theoretical models. 
This lack of resolution  was  inevitable because exploring 
large dimensional parameter space is 
extremely time-consuming even with the help of MCMC analysis.

Recently, on the other hand, two of the present authors
 (RN and JY) performed reconstruction of primordial
power spectrum from the angular power spectrum of the CMB temperature
fluctuations, $C_\ell^{TT}$, of the five-year WMAP data 
(hereafter WMAP5)\citep{WMAP5a,WMAP5} using two different 
non-parametric methods, 
namely, the cosmic inversion method \citep{MSY02,MSY03,KSY04,KSY05,
KMSY04,2008PhRvD..78l3002N} and 
the maximum-likelihood reconstruction 
method\citep{Tocchini,silk,2009PhRvD..79d3010N}.  They have 
probed the primordial power spectrum with finer resolution 
than aforementioned MCMC analyses and 
found 
an anomalously large 
deviation from the best-fit power-law primordial
spectrum of WMAP5
around the wavenumber $k\simeq 0.009\mathrm{Mpc}^{-1}$
or $kd\simeq 124$ where $d=1.43\times 10^4$Mpc is the 
distance to the last
 scattering surface.
This scale
corresponds to the multipole $\ell \simeq 120$ and the length 
scale $r\simeq 710$Mpc. 

In their reconstruction procedure, however, they had to fix values of the
cosmological parameters and adopt an {\it ad hoc} smoothing prior to
ensure sensible reconstruction.  The latter tends to discard
a large part of the information on the fine structure which may have
affected the significance of the claimed spectral feature itself.  
Furthermore although the reconstructed power spectrum exhibited
oscillatory features with both peaks and dips, the result of 
decomposition of the continuous reconstructed curve to statistically
independent band-powers,
which is necessary to discuss their statistical significance,
revealed only a $3.3\sigma$ peak but no dips \citep{2009PhRvD..79d3010N}.

The purpose of this paper is to focus on the fine structures of the 
spectral shape.  We use a forward analysis 
 implemented by MCMC simulation
which circumvents the above-mentioned
difficulties peculiar to the inverse mapping approach.
We also overcome the lack of resolution, which was a very serious
problem of previous MCMC analyses as mentioned above, by
searching for and 
concentrating on the most prominent localized feature 
imprinted on the otherwise power-law spectrum.
 
\section{Method}

First  we prepare various model power spectra
in which a localized spectral feature is imprinted on the power-law
with three fitting parameters.  We then calculate
 the temperature-temperature (TT) angular
power spectrum and the cross correlation
between the temperature and the E-mode polarization (TE)
to find the best-fit values of these parameters as well as those of
other cosmological parameters in the  $\Lambda$CDM model
using  the  CosmoMC\citep{MC} code.

We start with the following three
types of three-parameter
modulations. 
\begin{description}
\item[$\Lambda$-type:] a $\Lambda$-shaped
peak is introduced at $k=k_\ast$ and
connected to a power-law $A(k)=A(k/k_0)^{n_s-1}$
by straight lines in the $\log k-A(k)$
plane.  The location of the peak, $k_\ast$,
its width,  and height are additional fitting parameters. 
Here  $k_0\equiv 0.002{\rm Mpc}^{-1}$ is the pivot scale.
\item[$_{\rm V}~\!\!\!\!^\Lambda$-type:] a peak at $k=k_\ast$
and a dip at $k<k_\ast$  
with the same amplitude of deviation from the standard power-law
are connected by piecewise straight lines
to $A(k)=A(k/k_0)^{n_s-1}$ in the 
$\log k-A(k)$ plane. Again the height and the width of the peak are the
parameters characterizing modulation besides the location $k_\ast$.  
\item[$S$-type:] peaks and dips
are characterized by the following  smooth
function.
\beq
A(k)=A(k/k_0)^{n_s-1}+B (k/k_0)^{n_s-1}
 \exp[-(k-k_\ast)^2/\kappa^2]\cos[\pi(k-k_\ast)/\kappa],
\eeq
where $k_\ast$, $B$, and $\kappa$ are additional fitting parameters.
\end{description}
We assume flat priors for the additional parameters to cover
the parameter range as broad as possible.
  For the $S$-type modulation, for example, 
we take  $10^{10}B=[0,10^3]$ and $10^4\kappa=[1,10]\mathrm{Mpc}^{-1}$.
As for the locaiton of the feature, $k_\ast$, we started our
calculation allowing it to vary in the full range of 
observationally accessible domain with a sufficient accuracy 
so that we could detect a modulation
localized anywhere in $k$ space.  We have confirmed, 
however, that in the range of the wavenumber accessible by current
observation the
most prominent  feature is located at $k_\ast d\simeq 124$
as was found in the reconstruction analysis \citep{2009PhRvD..79d3010N}
together with other
possible modulations whose statistical significance is smaller.
In fact, if we allowed $k_\ast$ to vary beyond $k_\ast d>135$ and/or
$k_\ast d < 115$, other features would also contribute to 
MCMC analysis, albeit with little 
significance, in addition to the
most prominent one at $k_\ast d\simeq 124$.  Then our three-parameter
models would  no longer be
a good parametrization and the MCMC calculation would not converge
properly.  We therefore decided to limit the range of $k_\ast$ to
$k_\ast d =[115,135]$ in the final simulations in order to calculate
the probability distribution of the amplitude of the most prominent
modulation, $B$, properly.

In order to check the stability of our result 
with respect to other features localized at  different wavenumbers
we ran MCMC analysis incorporating
 another feature beyond $k_\ast d>135$ at
the same time with  three more fitting parameters.  As a result
we found that the inclusion of 
additional  parameters to incorporate another feature
does not affect  the posterior
distributions of 
the original three parameters for the most prominent feature at $k_\ast
d \simeq 124$, which implies that we can treat the relevant feature
independently from other possible features at different wavenumbers.

Based on the above observations we have also adopted another
model to fit the feature around the most prominent modulation
at $k_\ast d\simeq 124$ as follows.
\begin{description}
\item[$W$-type:]
three-step modulation on a power law with the functional form,
\begin{eqnarray}
A(k)=&& 
 A(k/k_0)^{n_s-1}+B_<\theta(kd-114)\theta(122-kd) \nonumber\\
&&
+B\,\theta(kd-122)\theta(126-kd)+B_>\theta(kd-126)\theta(134-kd),
\end{eqnarray}
where $B$, $B_<$, and $B_>$ are  fitting parameters.
\end{description}

Among the four models we use,
the $\Lambda$-type modulation contains
 only an extra power  so
that it enhances the dispersion of fluctuations compared with the
background power-law spectrum for the same value of $A$.  
In the other three types of models, both
excess and deficit are incorporated so that they do not necessarily
affect overall normalization of fluctuations.  In these three models, we
put $A(k)=0$ whenever $A(k)$ takes a negative value.  In the course of
MCMC calculations, we have found a degeneracy between the width and the
height of modulation in $\Lambda$- and $_{\rm V}~\!\!\!\!^\Lambda$-type
models and they tend to give a larger modulation amplitude with a
smaller width.  Such a tendency is undesirable from physical point of
view because $A(k)$ should be positive definite.  Hence we fixed the
width of modulation to $\Delta \ln k =0.043$ in these two models which
is the best-fit value in our first trial run and this choice is
consistent with the result of the $S$-type model where the width
$\kappa$ is treated as a free fitting parameter.  

\section{Results}

The result of MCMC calculations is summerized in Table 1.
First by incorporating an excess with the $\Lambda$-type model, the
effective $\chi^2_{\rm eff}\equiv -2\ln\CL$, 
where $\CL$ is the likelihood
function, improves by $\Delta\chi^2_{\rm eff}=-6.5$
 with three extra parameters
added to the standard power-law $\Lambda$CDM model.
To this end this model satisfies the AIC.
Inspecting the details of $\Delta\chi^2_{\rm eff}$, however,
we find that the improvement of $-6.5$
is entirely due to the better fit to the TE data.  Although the fit
to TT data improves around $\ell \approx 120$,
it gets worse for $\ell \lesssim 100$.
This is because the transfer function from $P(k)$ 
to  $C_\ell^{TT}$
is non-vanishing
for $\ell \lesssim kd $ so that an excessive power around 
$kd=k_\ast d$ increases all $C_\ell^{TT}$ for $\ell \lesssim 120$ 
to affect overall normalization in an unwanted manner.

Therefore even if it was not detected in a band-power 
analysis \citep{2009PhRvD..79d3010N},
we should take neighboring dips observed in the reconstructed curve 
as in the other three models, where $\chi^2_{\rm eff}$ improves as much
as $\Delta\chi^2_{\rm eff}=-16 \sim -22$ with the same numbers of extra
parameters. We can regard it  extremely significant and interpret that 
the result strongly suggests a sharp and strong oscillatory
deviation from a power-law around $k=k_\ast$. 
The marginalized 1D distributions of the parameter $B$ made by MCMC
calculations are shown in figure 1.  Based on the posterior statistical
distribution of the amplitude of spectral modulation, we find that
the pure power-law model with $B=0$ is $4.0\sigma$ ($4.1\sigma$)
away from the mean value of $B$ for $S$-type 
($_{\rm V}~\!\!\!\!^\Lambda$-type) model, respectively.

 \begin{figure}
 \includegraphics[width=0.5\textwidth]{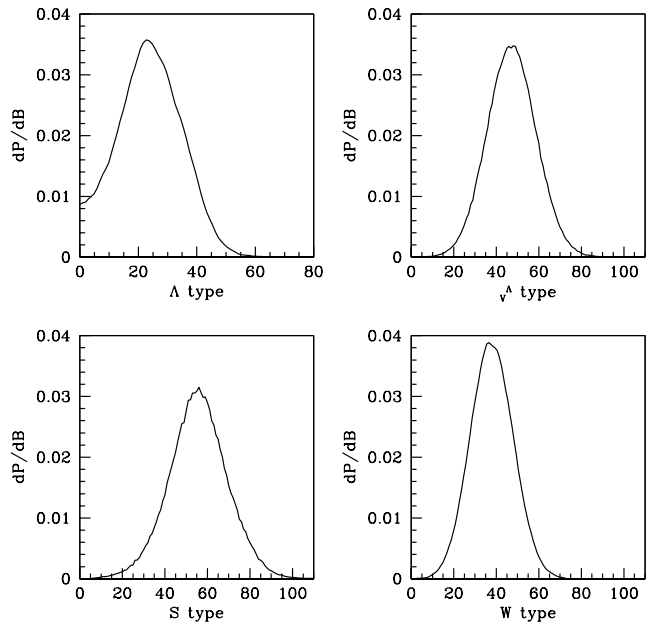}
 \caption{Posterior distribution functions for parameters of the
 modulation amplitude $B$.
 Here the horizontal axis denotes the parameter $B$. For all models the
 simple power law spectrum corresponds to $B=0$. 
}
 \end{figure}

We also examined the probability to find such a large deviation by
accumulating more samples to explore the tail of the posterior 
probability distribution for the $S$-type model
which improved $\chi^2_{\rm eff}$ the best among the models we adopted. 
We have found $304$ samples
in the smallest bin of the modulation parameter with $B < 10^{-10}
\simeq A/23$, from
$6279082$ MCMC samples generated with temperature parameter 
$T=1$ \citep{MacKayBook}. Hence the
relative probability is  
\begin{equation}
P(B<1\times 10^{-10})=4.8\times 10^{-5}~.
\end{equation}
In order to check whether we have explored the tails of the
distribution with sufficient accuracy,
we have ran simulations with different temperature parameters
($T=1.5$ ,$2$) in the MCMC analysis and confirmed that the probability
converges to $P= (3.9\times 10^{-5}, 5.6\times 10^{-5})$, 
respectively. Therefore we
conclude that the posterior probability for $B=0$ to be the case is only
${\cal O}(10^{-5})$.

We  note that among the improvement of $\Delta \chi^2_{\rm eff}=-22$
for the $S$-type spectrum,    $-14$ is due to the
improvement of the fit to the TT
power spectrum, $C_\ell^{TT}$, while
the remaining $-8$ is from
the TE cross correlation, $C_\ell^{TE}$.  
Figures 2 and 3 depict $C_\ell^{TT}$ and  $C_\ell^{TE}$
 in the relevant range. 
It is intriguing that significant portion of the improvement
of the likelihood comes from the TE data which was not used in the
reconstruction approach at all \citep{2009PhRvD..79d3010N}.  
Therefore this may be regarded as
an independent support to our discovery of a non-power-law feature 
in the primordial power spectrum.  

\begin{figure}
\includegraphics[width=0.5\textwidth]{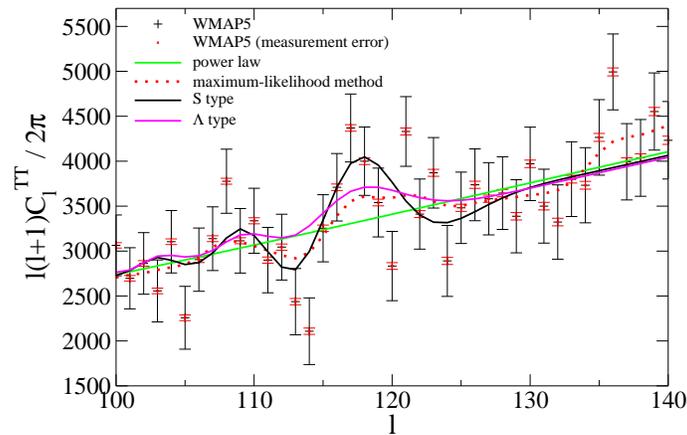}
\caption{Angular power spectrum of CMB temperature anisotropy
in the range $100 \leq \ell \leq 140$.
Data points are from WMAP5 with smaller error bars indicating only the
measurement errors.  Green line is the result of
best-fit power-law $\Lambda$CDM model, while the most wavy black curve
and the less wavy magenta curve represent the results of $S$-type
modulation and $\Lambda$-type modulation, respectively.  The dotted
curve is the angular power spectrum recalculated from $P(k)$ 
reconstructed by the maximum-likelihood reconstruction 
method \citep{2009PhRvD..79d3010N}.
}
\end{figure}

In order to pursue it, we have performed two distinct sets of MCMC
calculations
 one using only TT data and the other TE data.  In both cases the
$S$-type modulation was used and the cosmological parameters have been
fixed to the best-fit values for this model shown in Table 1.
To ensure the stability of calculation we discarded any $A(k)$ reaching
negative values in the calculation instead of setting $A(k)=0$.
As a result we have obtained the
following values of the model parameters from TT and TE data,
respectively. 
\begin{eqnarray}
 k_\ast^{TT}d=124.54^{+1.64}_{-1.42},~&~&~
k_\ast^{TE}d=123.95^{+3.87}_{-3.24}, \nonumber \\
10^4\kappa^{TT}=3.63^{+1.52}_{-1.26},~&~&~
10^4\kappa^{TE}=7.23^{+2.77}_{-3.13}, \nonumber \\
B^{TT}=\lmk 3.89^{+1.38}_{-2.26}\rmk\times 10^{-9},~&~&~
B^{TE}=\lmk 4.13^{+1.17}_{-2.21}\rmk\times 10^{-9}, \nonumber
\end{eqnarray}
where the errors represent 95\% upper and lower bounds.
Here the location of the peak is determined in  good shape in
both cases and the results are in good agreement with each other.
On the other hand, TE data fails to constrain the width parameter
$\kappa$ and its upper bound is essentially determined by the
prior cutoff.  Finally as for the amplitude of modulation $B$,
both datasets give similar posterior distributions and we find
the relative frequency to find vanishingly small modulation is
$\order(10^{-3})$ in both cases.  More precisely, we find
\[
 P(B^{TT}< 1\times 10^{-10})=2\times 10^{-3}~{\rm and}~
P(B^{TE}< 1\times 10^{-10})=1\times 10^{-3}. 
\]
 Thus we can conclude both TT data
and TE data suggest the existence of a feature at the same wavenumber
separately and mutually consistently.

\begin{figure}
\includegraphics[width=0.5\textwidth]{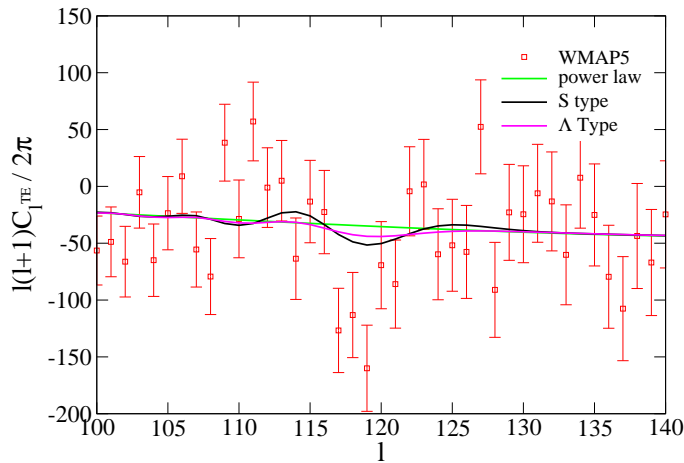}
\caption{Angular cross correlation of CMB temperature and E-mode
polarization in the same range.
Data points are the observational result of WMAP5.
Nearly straight green line is the result of
best-fit power-law $\Lambda$CDM model, while the black curve
and the magenta curve represent the results of $S$-type
modulation and $\Lambda$-type modulation, respectively.
}
\end{figure}

\if 
Let us next discuss whether the observed feature is simply a
realization of a rare event in the standard power-law primordial
spectrum or a result of some nontrivial physics which induces a
modulation in the power spectrum.  We have found an event with posterior
probability  $\sim 4.8\times 10^{-5}$  after searching in the range
$\Delta kd=20$.  Since we can probe the shape of the primordial spectrum
with sufficient accuracy in $40 \lesssim kd \lesssim 380$ with WMAP5 
\citep{2009PhRvD..79d3010N}, the probability to find one such a feature
in the entire observable domain would then be $\sim 8\times 10^{-4}$
if our finding is just a single realization of a rare event
which may take place at an arbitrary
wavenumber with uniform probability in the standard power-law model.
While this probability may not be small enough for us to reject the
simple power-law model, it is certainly sufficiently significant to 
motivate serious theoretical study to realize such an interesting
feature.
\fi
 
Finally we note that such
 a feature also shifts the best-fit values of other cosmological
parameters slightly as seen in Table 1.  
Although the shifts are smaller than the observational errors of 
five-year WMAP data, some of them are larger than the expected errors
in the Planck mission.
It has been already shown that globally non-power-law spectra generally
induce shifts in the estimated values of the cosmological parameters
compared with the standard power-law $\Lambda$CDM 
model \citep{1,4,4.2,6,7} because there is a degeneracy between 
the global shape of $P(k)$
and the cosmological parameters whose values are imprinted on the 
shape of the acoustic oscillation.  Nevertheless 
it is
intriguing that features localized in a narrow range of wavenumber
as discussed here also changes their values at a non-negligible
level for the next-generation analysis.

\begin{table}
\begin{tabular}{cccccccccc}
\hline\hline
& power law & $\Lambda$-type & $_{\rm V}~\!\!\!^\Lambda$-type
& $S$-type~ & $W$-type~ & $\Delta_{\max}$ & WMAP5 & Planck \\
\hline
$\Omega_b $ 
& 0.0438& 0.0441 & 0.0443 & 0.0441 & 0.0444 &  0.0006 & 0.0030 & 0.0003\\
$\Omega_m $ 
&0.256 & 0.256 & 0.260 & 0.257 & 0.262 & 0.006 & 0.027 \\
$\Omega_\Lambda$ 
&0.744 & 0.744 & 0.740 & 0.743 & 0.738 &  0.006 & 0.015& 0.009\\
$H_0$ 
&72.1 & 72.1 & 71.7 &  72.0 & 71.6 & 0.5 & 2.7&2.7\\
$10^{10}A$ 
& 23.88 & 23.24 & 23.51 & 23.34 & 23.90 & 0.54 & 1.12 \\
$n_s$ 
&0.964 & 0.975 & 0.969 &  0.970 & 0.964 & 0.006 & 0.015 & 0.0045 \\
%
%
%
$\tau$ 
&0.0864 & 0.0879 & 0.0846 & 0.0835  & 0.0845 & 0.0029 & 0.017 & 0.005\\
%
$\Delta \chi^2_{\rm eff}$ 
& 0 & $-6.5$ & $-19$ & $-22$ & $-16$ &  \\
$k_\ast d$  
&& 124.5 & 124.4 & 124.5 & $-$ & \\
$10^{10}\rr$ 
&& 23.80 & 47.26 & 55.66 & 37.95 
& \\
\hline
\end{tabular}
\caption{Cosmological parameters estimated from CosmoMC code using four
 types of modulated power spectra.  Here $\Omega_b, \Omega_m$ and
 $\Omega_\Lambda$ are fractions of cosmic energy density in baryons,
 matters and cosmological constant, respectively, $H_0$ is the Hubble
 parameter in unit of km/s/Mpc, and $\tau$ is the optical depth to the
 last scattering surface.  For $\Lambda$- and $_{\rm
 V}~\!\!\!^\Lambda$-type models the parameter $B$ stands for the
 difference between the peak amplitude and the power-law background value
 at $k=k_\ast$.  $\Delta\chi^2_{\rm eff}$ represents improvement of
 goodness-of-fit $\chi^2_{\rm eff}$ for the best-fit model of each
 modulated spectrum from the most-updated best-fit power-law $\Lambda$CDM
 model.  $\Delta_{\max}$ is the maximum of the difference of each
 parameter between the power-law model and models with modulated spectra
 except for the $\Lambda$-type which does not improve the fit.  Column
 WMAP5 (Planck) represents errors observed (expected) by five-year WMAP
 (Planck).  Errors expected in Planck mission are taken from the ``Blue
 book''. 
 The mean value of $\kappa$ in
 the $S$-type model is $\kappa=3.58\times 10^{-4}$.} \label{tab:cosmo}
\end{table}

\section{Conclusion}

In this paper we have
found a spectral feature with a 
4$\sigma$ deviation from the best-fit power-law 
primordial power spectrum in
a narrow range around $k \simeq 0.009{\rm Mpc}^{-1}$, whose
existence is suggested separately by both TT data and TE data from
five-year WMAP data.  This feature can be
further tested and hopefully confirmed by forthcoming EE data from
the Planck mission.  

As emphasized above, the detected feature consists not only of a 
peak but also a dip.  Since we cannot lower the amplitude of
the power spectrum by any sources uncorrelated with the preexistent
fluctuations, this means that the feature was created together with
the bulk of the power-law spectrum, suggesting some nontrivial
phenomenon during inflation.  At the moment, while a number of 
inflation models have been proposed to produce spectral modulation,
we are not aware of any model that can realize a spectrum with 
such a localized feature.  Hence efforts should be made toward
construction of models that can account for the detected fine structures.

Currently inflation models are observationally constrained 
in terms of a small number of discrete parameters such as $n_s$,
$dn_s/d\ln k$, the tensor-to-scalar ratio $r$, and a nonlinearity
parameter $f_{NL}$. It is difficult to determine the 
underlying particle-physics model with these 
parameters alone, because there are huge degeneracies.
If our finding is further confirmed, it can certainly serve as an important
clue to single out the correct model.
On the other hand, whether this feature is a result of some nontrivial
physics or a realization of an extremely rare event, we must take
 it into
account in cosmological parameter estimation in the analysis of
forthcoming higher precision data.


\acknowledgements

We are grateful to Fran\a{c}cois Bouchet, Eiichiro Komatsu,  
David Spergel, and Masahiro Takada for useful communications.
This work was partially supported by 
JSPS Grant-in-Aid for Scientific
Research No.\ 19340054(JY) and 21740177(KI), the Grant-in-Aid for Scientific Research on
Innovative Areas No.\ 21111006(JY \& RN), 
JSPS Core-to-Core program  ``International Research Network on Dark
Energy'', and the Global Center of Excellence program at
Nagoya University "Quest for Fundamental Principles in the Universe:
from Particles to the Solar System and the Cosmos" from the MEXT of
Japan.

\bibliography{discord}

\begin{thebibliography}{30}
\expandafter\ifx\csname natexlab\endcsname\relax\def\natexlab#1{#1}\fi
\expandafter\ifx\csname bibnamefont\endcsname\relax
  \def\bibnamefont#1{#1}\fi
\expandafter\ifx\csname bibfnamefont\endcsname\relax
  \def\bibfnamefont#1{#1}\fi
\expandafter\ifx\csname citenamefont\endcsname\relax
  \def\citenamefont#1{#1}\fi
\expandafter\ifx\csname url\endcsname\relax
  \def\url#1{\texttt{#1}}\fi
\expandafter\ifx\csname urlprefix\endcsname\relax\def\urlprefix{URL }\fi
\providecommand{\bibinfo}[2]{#2}
\providecommand{\eprint}[2][]{\url{#2}}

\bibitem[{\citenamefont{{Guth}}(1981)}]{guth}
\bibinfo{author}{\bibfnamefont{A.~H.} \bibnamefont{{Guth}}},
  \bibinfo{journal}{\prd} \textbf{\bibinfo{volume}{23}}, \bibinfo{pages}{347}
  (\bibinfo{year}{1981}).

\bibitem[{\citenamefont{{Sato}}(1981)}]{sato}
\bibinfo{author}{\bibfnamefont{K.}~\bibnamefont{{Sato}}},
  \bibinfo{journal}{\mnras} \textbf{\bibinfo{volume}{195}},
  \bibinfo{pages}{467} (\bibinfo{year}{1981}).

\bibitem[{\citenamefont{{Starobinsky}}(1980)}]{staro}
\bibinfo{author}{\bibfnamefont{A.~A.} \bibnamefont{{Starobinsky}}},
  \bibinfo{journal}{Phys. Lett. B} \textbf{\bibinfo{volume}{91}},
  \bibinfo{pages}{99} (\bibinfo{year}{1980}).

\bibitem[{\citenamefont{{Linde}}(2008)}]{infreview}
\bibinfo{author}{\bibfnamefont{A.}~\bibnamefont{{Linde}}}, in
  \emph{\bibinfo{booktitle}{Lecture Notes in Physics, Berlin Springer Verlag}},
  edited by \bibinfo{editor}{\bibfnamefont{M.}~\bibnamefont{{Lemoine}}},
  \bibinfo{editor}{\bibfnamefont{J.}~\bibnamefont{{Martin}}}, \bibnamefont{and}
  \bibinfo{editor}{\bibfnamefont{P.}~\bibnamefont{{Peter}}}
  (\bibinfo{year}{2008}), vol. \bibinfo{volume}{738} of
  \emph{\bibinfo{series}{Lecture Notes in Physics, Berlin Springer Verlag}},
  pp. \bibinfo{pages}{1--+}.

\bibitem[{\citenamefont{{Hawking}}(1982)}]{yuragi}
\bibinfo{author}{\bibfnamefont{S.~W.} \bibnamefont{{Hawking}}},
  \bibinfo{journal}{Phys. Lett. B} \textbf{\bibinfo{volume}{115}},
  \bibinfo{pages}{295} (\bibinfo{year}{1982}).

\bibitem[{\citenamefont{{Starobinsky}}(1982)}]{yuragi2}
\bibinfo{author}{\bibfnamefont{A.~A.} \bibnamefont{{Starobinsky}}},
  \bibinfo{journal}{Phys. Lett. B} \textbf{\bibinfo{volume}{117}},
  \bibinfo{pages}{175} (\bibinfo{year}{1982}).

\bibitem[{\citenamefont{{Guth} and {Pi}}(1982)}]{yuragi3}
\bibinfo{author}{\bibfnamefont{A.~H.} \bibnamefont{{Guth}}} \bibnamefont{and}
  \bibinfo{author}{\bibfnamefont{S.-Y.} \bibnamefont{{Pi}}},
  \bibinfo{journal}{\prl} \textbf{\bibinfo{volume}{49}}, \bibinfo{pages}{1110}
  (\bibinfo{year}{1982}).

\bibitem[{\citenamefont{{Kawasaki} et~al.}(2003)\citenamefont{{Kawasaki},
  {Yamaguchi}, and {Yokoyama}}}]{Kawasaki:2003zv}
\bibinfo{author}{\bibfnamefont{M.}~\bibnamefont{{Kawasaki}}},
  \bibinfo{author}{\bibfnamefont{M.}~\bibnamefont{{Yamaguchi}}},
  \bibnamefont{and}
  \bibinfo{author}{\bibfnamefont{J.}~\bibnamefont{{Yokoyama}}},
  \bibinfo{journal}{\prd} \textbf{\bibinfo{volume}{68}},
  \bibinfo{pages}{023508} (\bibinfo{year}{2003}),
  \eprint{arXiv:hep-ph/0304161}.

\bibitem[{\citenamefont{{Covi} et~al.}(2006)\citenamefont{{Covi}, {Hamann},
  {Melchiorri}, {Slosar}, and {Sorbera}}}]{4}
\bibinfo{author}{\bibfnamefont{L.}~\bibnamefont{{Covi}}},
  \bibinfo{author}{\bibfnamefont{J.}~\bibnamefont{{Hamann}}},
  \bibinfo{author}{\bibfnamefont{A.}~\bibnamefont{{Melchiorri}}},
  \bibinfo{author}{\bibfnamefont{A.}~\bibnamefont{{Slosar}}}, \bibnamefont{and}
  \bibinfo{author}{\bibfnamefont{I.}~\bibnamefont{{Sorbera}}},
  \bibinfo{journal}{\prd} \textbf{\bibinfo{volume}{74}},
  \bibinfo{pages}{083509} (\bibinfo{year}{2006}),
  \eprint{arXiv:astro-ph/0606452}.

\bibitem[{\citenamefont{{Bridges} et~al.}(2006)\citenamefont{{Bridges},
  {Lasenby}, and {Hobson}}}]{4.2}
\bibinfo{author}{\bibfnamefont{M.}~\bibnamefont{{Bridges}}},
  \bibinfo{author}{\bibfnamefont{A.~N.} \bibnamefont{{Lasenby}}},
  \bibnamefont{and} \bibinfo{author}{\bibfnamefont{M.~P.}
  \bibnamefont{{Hobson}}}, \bibinfo{journal}{\mnras}
  \textbf{\bibinfo{volume}{369}}, \bibinfo{pages}{1123} (\bibinfo{year}{2006}),
  \eprint{arXiv:astro-ph/0511573}.

\bibitem[{\citenamefont{{Martin} and {Ringeval}}(2006)}]{5}
\bibinfo{author}{\bibfnamefont{J.}~\bibnamefont{{Martin}}} \bibnamefont{and}
  \bibinfo{author}{\bibfnamefont{C.}~\bibnamefont{{Ringeval}}},
  \bibinfo{journal}{\jcap} \textbf{\bibinfo{volume}{8}}, \bibinfo{pages}{9}
  (\bibinfo{year}{2006}), \eprint{arXiv:astro-ph/0605367}.

\bibitem[{\citenamefont{{Hunt} and {Sarkar}}(2007)}]{7}
\bibinfo{author}{\bibfnamefont{P.}~\bibnamefont{{Hunt}}} \bibnamefont{and}
  \bibinfo{author}{\bibfnamefont{S.}~\bibnamefont{{Sarkar}}},
  \bibinfo{journal}{\prd} \textbf{\bibinfo{volume}{76}},
  \bibinfo{pages}{123504} (\bibinfo{year}{2007}), \eprint{0706.2443}.

\bibitem[{\citenamefont{{Spergel} et~al.}(2007)\citenamefont{{Spergel}, {Bean},
  {Dor{\'e}}, {Nolta}, {Bennett}, {Dunkley}, {Hinshaw}, {Jarosik}, {Komatsu},
  {Page} et~al.}}]{WMAP3COSMO}
\bibinfo{author}{\bibfnamefont{D.~N.} \bibnamefont{{Spergel}}},
  \bibinfo{author}{\bibfnamefont{R.}~\bibnamefont{{Bean}}},
  \bibinfo{author}{\bibfnamefont{O.}~\bibnamefont{{Dor{\'e}}}},
  \bibinfo{author}{\bibfnamefont{M.~R.} \bibnamefont{{Nolta}}},
  \bibinfo{author}{\bibfnamefont{C.~L.} \bibnamefont{{Bennett}}},
  \bibinfo{author}{\bibfnamefont{J.}~\bibnamefont{{Dunkley}}},
  \bibinfo{author}{\bibfnamefont{G.}~\bibnamefont{{Hinshaw}}},
  \bibinfo{author}{\bibfnamefont{N.}~\bibnamefont{{Jarosik}}},
  \bibinfo{author}{\bibfnamefont{E.}~\bibnamefont{{Komatsu}}},
  \bibinfo{author}{\bibfnamefont{L.}~\bibnamefont{{Page}}},
  \bibnamefont{et~al.}, \bibinfo{journal}{\apjs}
  \textbf{\bibinfo{volume}{170}}, \bibinfo{pages}{377} (\bibinfo{year}{2007}),
  \eprint{arXiv:astro-ph/0603449}.

\bibitem[{\citenamefont{{Ichiki} and {Nagata}}(2009)}]{2009PhRvD..80h3002I}
\bibinfo{author}{\bibfnamefont{K.}~\bibnamefont{{Ichiki}}} \bibnamefont{and}
  \bibinfo{author}{\bibfnamefont{R.}~\bibnamefont{{Nagata}}},
  \bibinfo{journal}{\prd} \textbf{\bibinfo{volume}{80}},
  \bibinfo{pages}{083002} (\bibinfo{year}{2009}).

\bibitem[{\citenamefont{{Akaike}}(1974)}]{akaike}
\bibinfo{author}{\bibfnamefont{H.}~\bibnamefont{{Akaike}}},
  \bibinfo{journal}{IEEE Trans. Auto. Control} \textbf{\bibinfo{volume}{19}},
  \bibinfo{pages}{716} (\bibinfo{year}{1974}).

\bibitem[{\citenamefont{{Hinshaw} et~al.}(2009)\citenamefont{{Hinshaw},
  {Weiland}, {Hill}, {Odegard}, {Larson}, {Bennett}, {Dunkley}, {Gold},
  {Greason}, {Jarosik} et~al.}}]{WMAP5a}
\bibinfo{author}{\bibfnamefont{G.}~\bibnamefont{{Hinshaw}}},
  \bibinfo{author}{\bibfnamefont{J.~L.} \bibnamefont{{Weiland}}},
  \bibinfo{author}{\bibfnamefont{R.~S.} \bibnamefont{{Hill}}},
  \bibinfo{author}{\bibfnamefont{N.}~\bibnamefont{{Odegard}}},
  \bibinfo{author}{\bibfnamefont{D.}~\bibnamefont{{Larson}}},
  \bibinfo{author}{\bibfnamefont{C.~L.} \bibnamefont{{Bennett}}},
  \bibinfo{author}{\bibfnamefont{J.}~\bibnamefont{{Dunkley}}},
  \bibinfo{author}{\bibfnamefont{B.}~\bibnamefont{{Gold}}},
  \bibinfo{author}{\bibfnamefont{M.~R.} \bibnamefont{{Greason}}},
  \bibinfo{author}{\bibfnamefont{N.}~\bibnamefont{{Jarosik}}},
  \bibnamefont{et~al.}, \bibinfo{journal}{\apjs}
  \textbf{\bibinfo{volume}{180}}, \bibinfo{pages}{225} (\bibinfo{year}{2009}),
  \eprint{0803.0732}.

\bibitem[{\citenamefont{{Komatsu} et~al.}(2009)\citenamefont{{Komatsu},
  {Dunkley}, {Nolta}, {Bennett}, {Gold}, {Hinshaw}, {Jarosik}, {Larson},
  {Limon}, {Page} et~al.}}]{WMAP5}
\bibinfo{author}{\bibfnamefont{E.}~\bibnamefont{{Komatsu}}},
  \bibinfo{author}{\bibfnamefont{J.}~\bibnamefont{{Dunkley}}},
  \bibinfo{author}{\bibfnamefont{M.~R.} \bibnamefont{{Nolta}}},
  \bibinfo{author}{\bibfnamefont{C.~L.} \bibnamefont{{Bennett}}},
  \bibinfo{author}{\bibfnamefont{B.}~\bibnamefont{{Gold}}},
  \bibinfo{author}{\bibfnamefont{G.}~\bibnamefont{{Hinshaw}}},
  \bibinfo{author}{\bibfnamefont{N.}~\bibnamefont{{Jarosik}}},
  \bibinfo{author}{\bibfnamefont{D.}~\bibnamefont{{Larson}}},
  \bibinfo{author}{\bibfnamefont{M.}~\bibnamefont{{Limon}}},
  \bibinfo{author}{\bibfnamefont{L.}~\bibnamefont{{Page}}},
  \bibnamefont{et~al.}, \bibinfo{journal}{\apjs}
  \textbf{\bibinfo{volume}{180}}, \bibinfo{pages}{330} (\bibinfo{year}{2009}),
  \eprint{0803.0547}.

\bibitem[{\citenamefont{{Matsumiya} et~al.}(2002)\citenamefont{{Matsumiya},
  {Sasaki}, and {Yokoyama}}}]{MSY02}
\bibinfo{author}{\bibfnamefont{M.}~\bibnamefont{{Matsumiya}}},
  \bibinfo{author}{\bibfnamefont{M.}~\bibnamefont{{Sasaki}}}, \bibnamefont{and}
  \bibinfo{author}{\bibfnamefont{J.}~\bibnamefont{{Yokoyama}}},
  \bibinfo{journal}{\prd} \textbf{\bibinfo{volume}{65}},
  \bibinfo{pages}{083007} (\bibinfo{year}{2002}),
  \eprint{arXiv:astro-ph/0111549}.

\bibitem[{\citenamefont{{Matsumiya} et~al.}(2003)\citenamefont{{Matsumiya},
  {Sasaki}, and {Yokoyama}}}]{MSY03}
\bibinfo{author}{\bibfnamefont{M.}~\bibnamefont{{Matsumiya}}},
  \bibinfo{author}{\bibfnamefont{M.}~\bibnamefont{{Sasaki}}}, \bibnamefont{and}
  \bibinfo{author}{\bibfnamefont{J.}~\bibnamefont{{Yokoyama}}},
  \bibinfo{journal}{\jcap} \textbf{\bibinfo{volume}{2}}, \bibinfo{pages}{3}
  (\bibinfo{year}{2003}), \eprint{arXiv:astro-ph/0210365}.

\bibitem[{\citenamefont{{Kogo} et~al.}(2004{\natexlab{a}})\citenamefont{{Kogo},
  {Sasaki}, and {Yokoyama}}}]{KSY04}
\bibinfo{author}{\bibfnamefont{N.}~\bibnamefont{{Kogo}}},
  \bibinfo{author}{\bibfnamefont{M.}~\bibnamefont{{Sasaki}}}, \bibnamefont{and}
  \bibinfo{author}{\bibfnamefont{J.}~\bibnamefont{{Yokoyama}}},
  \bibinfo{journal}{\prd} \textbf{\bibinfo{volume}{70}},
  \bibinfo{pages}{103001} (\bibinfo{year}{2004}{\natexlab{a}}),
  \eprint{arXiv:astro-ph/0409052}.

\bibitem[{\citenamefont{{Kogo} et~al.}(2005)\citenamefont{{Kogo}, {Sasaki}, and
  {Yokoyama}}}]{KSY05}
\bibinfo{author}{\bibfnamefont{N.}~\bibnamefont{{Kogo}}},
  \bibinfo{author}{\bibfnamefont{M.}~\bibnamefont{{Sasaki}}}, \bibnamefont{and}
  \bibinfo{author}{\bibfnamefont{J.}~\bibnamefont{{Yokoyama}}},
  \bibinfo{journal}{Prog. Theor. Phys.} \textbf{\bibinfo{volume}{114}},
  \bibinfo{pages}{555} (\bibinfo{year}{2005}), \eprint{arXiv:astro-ph/0504471}.

\bibitem[{\citenamefont{{Kogo} et~al.}(2004{\natexlab{b}})\citenamefont{{Kogo},
  {Matsumiya}, {Sasaki}, and {Yokoyama}}}]{KMSY04}
\bibinfo{author}{\bibfnamefont{N.}~\bibnamefont{{Kogo}}},
  \bibinfo{author}{\bibfnamefont{M.}~\bibnamefont{{Matsumiya}}},
  \bibinfo{author}{\bibfnamefont{M.}~\bibnamefont{{Sasaki}}}, \bibnamefont{and}
  \bibinfo{author}{\bibfnamefont{J.}~\bibnamefont{{Yokoyama}}},
  \bibinfo{journal}{\apj} \textbf{\bibinfo{volume}{607}}, \bibinfo{pages}{32}
  (\bibinfo{year}{2004}{\natexlab{b}}), \eprint{arXiv:astro-ph/0309662}.

\bibitem[{\citenamefont{{Nagata} and {Yokoyama}}(2008)}]{2008PhRvD..78l3002N}
\bibinfo{author}{\bibfnamefont{R.}~\bibnamefont{{Nagata}}} \bibnamefont{and}
  \bibinfo{author}{\bibfnamefont{J.}~\bibnamefont{{Yokoyama}}},
  \bibinfo{journal}{\prd} \textbf{\bibinfo{volume}{78}},
  \bibinfo{pages}{123002} (\bibinfo{year}{2008}), \eprint{0809.4537}.

\bibitem[{\citenamefont{{Tocchini-Valentini}
  et~al.}(2005)\citenamefont{{Tocchini-Valentini}, {Douspis}, and
  {Silk}}}]{Tocchini}
\bibinfo{author}{\bibfnamefont{D.}~\bibnamefont{{Tocchini-Valentini}}},
  \bibinfo{author}{\bibfnamefont{M.}~\bibnamefont{{Douspis}}},
  \bibnamefont{and} \bibinfo{author}{\bibfnamefont{J.}~\bibnamefont{{Silk}}},
  \bibinfo{journal}{\mnras} \textbf{\bibinfo{volume}{359}}, \bibinfo{pages}{31}
  (\bibinfo{year}{2005}), \eprint{arXiv:astro-ph/0402583}.

\bibitem[{\citenamefont{{Tocchini-Valentini}
  et~al.}(2006)\citenamefont{{Tocchini-Valentini}, {Hoffman}, and
  {Silk}}}]{silk}
\bibinfo{author}{\bibfnamefont{D.}~\bibnamefont{{Tocchini-Valentini}}},
  \bibinfo{author}{\bibfnamefont{Y.}~\bibnamefont{{Hoffman}}},
  \bibnamefont{and} \bibinfo{author}{\bibfnamefont{J.}~\bibnamefont{{Silk}}},
  \bibinfo{journal}{\mnras} \textbf{\bibinfo{volume}{367}},
  \bibinfo{pages}{1095} (\bibinfo{year}{2006}),
  \eprint{arXiv:astro-ph/0509478}.

\bibitem[{\citenamefont{{Nagata} and {Yokoyama}}(2009)}]{2009PhRvD..79d3010N}
\bibinfo{author}{\bibfnamefont{R.}~\bibnamefont{{Nagata}}} \bibnamefont{and}
  \bibinfo{author}{\bibfnamefont{J.}~\bibnamefont{{Yokoyama}}},
  \bibinfo{journal}{\prd} \textbf{\bibinfo{volume}{79}},
  \bibinfo{pages}{043010} (\bibinfo{year}{2009}), \eprint{0812.4585}.

\bibitem[{\citenamefont{{Lewis} and {Bridle}}(2002)}]{MC}
\bibinfo{author}{\bibfnamefont{A.}~\bibnamefont{{Lewis}}} \bibnamefont{and}
  \bibinfo{author}{\bibfnamefont{S.}~\bibnamefont{{Bridle}}},
  \bibinfo{journal}{\prd} \textbf{\bibinfo{volume}{66}},
  \bibinfo{pages}{103511} (\bibinfo{year}{2002}),
  \eprint{arXiv:astro-ph/0205436}.

\bibitem[{\citenamefont{MacKay}(2003)}]{MacKayBook}
\bibinfo{author}{\bibfnamefont{D.~J.~C.} \bibnamefont{MacKay}},
  \emph{\bibinfo{title}{Information Theory, Inference and Learning Algorithms}}
  (\bibinfo{publisher}{Cambrdige University Press}, \bibinfo{year}{2003}), ISBN
  \bibinfo{isbn}{0521642981}.

\bibitem[{\citenamefont{{Bridle} et~al.}(2003)\citenamefont{{Bridle}, {Lewis},
  {Weller}, and {Efstathiou}}}]{1}
\bibinfo{author}{\bibfnamefont{S.~L.} \bibnamefont{{Bridle}}},
  \bibinfo{author}{\bibfnamefont{A.~M.} \bibnamefont{{Lewis}}},
  \bibinfo{author}{\bibfnamefont{J.}~\bibnamefont{{Weller}}}, \bibnamefont{and}
  \bibinfo{author}{\bibfnamefont{G.}~\bibnamefont{{Efstathiou}}},
  \bibinfo{journal}{\mnras} \textbf{\bibinfo{volume}{342}},
  \bibinfo{pages}{L72} (\bibinfo{year}{2003}), \eprint{arXiv:astro-ph/0302306}.

\bibitem[{\citenamefont{{Shafieloo} et~al.}(2007)\citenamefont{{Shafieloo},
  {Souradeep}, {Manimaran}, {Panigrahi}, and {Rangarajan}}}]{6}
\bibinfo{author}{\bibfnamefont{A.}~\bibnamefont{{Shafieloo}}},
  \bibinfo{author}{\bibfnamefont{T.}~\bibnamefont{{Souradeep}}},
  \bibinfo{author}{\bibfnamefont{P.}~\bibnamefont{{Manimaran}}},
  \bibinfo{author}{\bibfnamefont{P.~K.} \bibnamefont{{Panigrahi}}},
  \bibnamefont{and}
  \bibinfo{author}{\bibfnamefont{R.}~\bibnamefont{{Rangarajan}}},
  \bibinfo{journal}{\prd} \textbf{\bibinfo{volume}{75}},
  \bibinfo{pages}{123502} (\bibinfo{year}{2007}),
  \eprint{arXiv:astro-ph/0611352}.

\end{thebibliography}


\newpage

\end{document}